\documentclass[11pt]{article}
\pdfoutput=1
\usepackage[utf8]{inputenc}
\usepackage{dcolumn}
\usepackage{bm}
\usepackage{graphicx}
\usepackage{diagbox}
\usepackage{multirow}
\usepackage{amssymb,amsmath}
\usepackage{multirow}
\usepackage{cite,color,url}
\usepackage[colorlinks=true
,urlcolor=blue
,anchorcolor=blue
,citecolor=blue
,filecolor=blue
,linkcolor=blue
,menucolor=blue
,linktocpage=true
,pdfproducer=medialab
,pdfa=true
]{hyperref}

\usepackage{slashed}
\usepackage{epsfig,psfrag,rotating,soul}
\usepackage{rotfloat}
\usepackage[font={small}]{caption}
\usepackage{colortbl}

\evensidemargin \oddsidemargin
\allowdisplaybreaks

\let\OLDthebibliography\thebibliography
\renewcommand\thebibliography[1]{
  \OLDthebibliography{#1}
  \setlength{\parskip}{0pt}
  \setlength{\itemsep}{0pt plus 0.3ex}
}

\definecolor{mygray}{gray}{0.85} 
\definecolor{myblue}{cmyk}{0.65, 0.37, 0.0, 0.19}

\begin{document}
\thispagestyle{empty}

\def\thefootnote{\fnsymbol{footnote}}

\begin{flushright}
IFT-UAM/CSIC-22-61
\end{flushright}

\vspace*{1cm}

\begin{center}

\begin{Large}
\textbf{\textsc{LHC search strategy for squarks in higgsino-LSP scenarios \\[.25em] with leptons and $b$-jets in the final state}}
\end{Large}

\vspace{1cm}

{\sc
Ernesto~Arganda$^{1, 2}$%
\footnote{{\tt \href{mailto:ernesto.arganda@csic.es}{ernesto.arganda@csic.es}}}%
, Antonio~Delgado$^{3}$%
\footnote{{\tt \href{mailto:adelgad2@nd.edu}{adelgad2@nd.edu}}}%
, Roberto~A.~Morales$^{2}$%
\footnote{{\tt \href{mailto:roberto.morales@fisica.unlp.edu.ar}{roberto.morales@fisica.unlp.edu.ar}}}%
,		 Mariano~Quir\'os$^{4}$%
\footnote{{\tt \href{quiros@ifae.es}{quiros@ifae.es}}}%

}

\vspace*{.7cm}

{\sl
$^1$Instituto de F\'{\i}sica Te\'orica UAM-CSIC, \\
C/ Nicol\'as Cabrera 13-15, Campus de Cantoblanco, 28049, Madrid, Spain

\vspace*{0.1cm}

$^2$IFLP, CONICET - Dpto. de F\'{\i}sica, Universidad Nacional de La Plata, \\ 
C.C. 67, 1900 La Plata, Argentina

\vspace*{0.1cm}

$^3$Department of Physics, University of Notre Dame, 225 Nieuwland Hall \\
Notre Dame, IN 46556, USA

\vspace*{0.1cm}

$^4$Institut de F\'{\i}sica d'Altes Energies (IFAE) and BIST, Campus UAB \\
08193, Bellaterra, Barcelona, Spain

}

\end{center}

\vspace{0.1cm}

\begin{abstract}
\noindent

The higgsino Lightest Supersymmetric Particle (LSP) scenario opens up the possibility of decays of strongly produced particles to an intermediate neutralino, due to the Yukawa-suppressed direct decays to the higgsino. Those decays produce multijet signals with a Higgs or a $Z$ boson being produced in the decay of the intermediate neutralino to the LSP. In this paper we study the discovery prospects of squarks that produce $b$-jets and leptons in the final state. Our collider analysis provides signal significances at the 3$\sigma$ level for luminosities of 1 ab$^{-1}$, and at the 5$\sigma$ level if we project these results for 3 ab$^{-1}$.

\end{abstract}

\def\thefootnote{\arabic{footnote}}
\setcounter{page}{0}
\setcounter{footnote}{0}

\newpage

\section{Introduction}
\label{intro}
The hierarchy problem and the existence of dark matter (DM) are two of the strongest motivations to enlarge the Standard Model (SM) with low energy R-parity conserving supersymmetry (MSSM)~\cite{Fayet:1976et,Fayet:1977yc,Nilles:1983ge,Haber:1984rc,Gunion:1984yn}. One of the consequences is the stability of the LSP, and hence a possible DM candidate. One appealing possibility to avoid the constrains coming from direct detection experiments is to assume that the higgsino is the LSP~\cite{Kowalska:2018toh}, and its implementation in the MSSM as done in Ref.~\cite{Delgado:2020url}.

Building on the intuition and knowledge gained in our previous works~\cite{Arganda:2021lpg,Arganda:2021iyr,Arganda:2021ajy} on the LHC phenomenology of MSSM scenarios with higgsino LSP, and taking into account the large number of events that we could expect at 14 TeV and 300 fb$^{-1}$ for the production of squark pairs~\cite{Arganda:2021ajy}, we consider in this work that the bino-like neutralino decays into the higgsino LSP plus a leptonic $Z$ boson as another interesting decay channel. This feature implies, on the one hand, the reduction of expected signal events, but however, on the other hand, it will provide a better control of all backgrounds, especially interesting to be able to discard the QCD multijet background.

This paper is organized as follows: in Section~\ref{collider} we present the details of the event simulation and develop our search strategy by means of the characterization of the signal against the background, while Section~\ref{results} is devoted to the discussion of our main results and a summary of the most important conclusions.

\section{Simulation and Collider Analysis}
\label{collider}

The signal and backgrounds were generated with {\tt MadGraph\_aMC@NLO 2.8.1}~\cite{Alwall:2014hca} for a center-of-mass energy of $\sqrt{s}$ = 14 TeV. The events are showered and hadronized using {\tt PYTHIA 8.2}~\cite{Sjostrand:2014zea}, and the detector effects are implemented with {\tt Delphes 3.3.3}~\cite{deFavereau:2013fsa}. We consider a working point for the efficiency of $b$-tagging of 0.75, with a rate of misidentification of $0.01$ for light jets and $0.1$ for $c$-jets. The internal analysis codes and simulation input files are available upon request to authors.

\begin{figure}[ht!]
	\begin{center}
		\begin{tabular}{c}
			\centering
			\hspace*{-3mm}
			\includegraphics[scale=0.75]{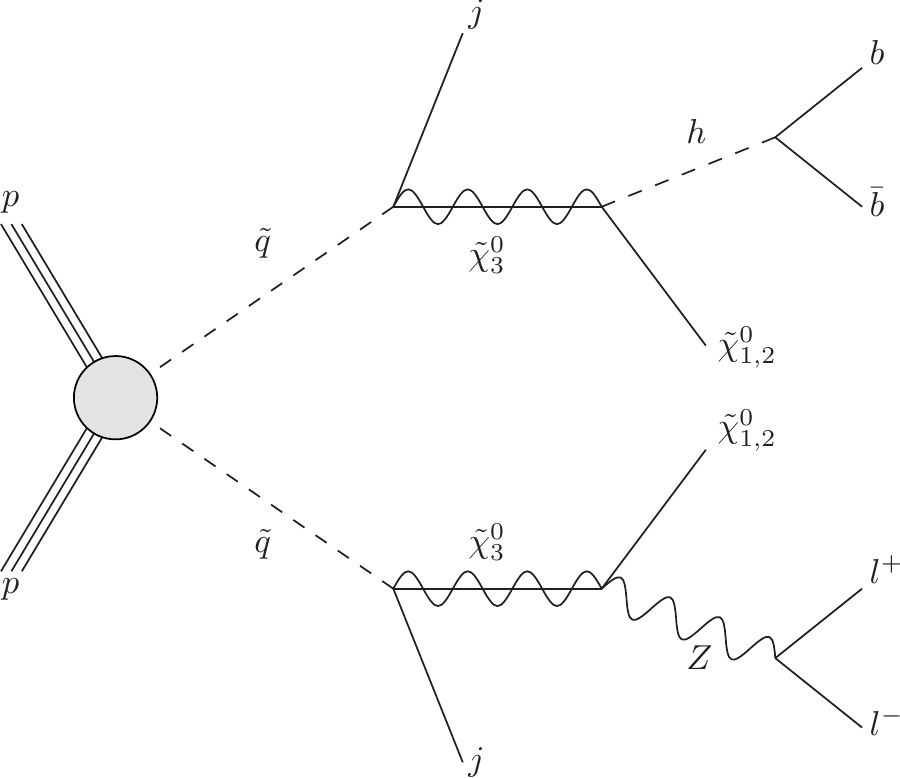} 
		\end{tabular}
		\caption{\it Representative Feynman diagram corresponding to our signature with $2 l+ 2 b + 2 j + E_T^\text{miss}$ in the final state.}
		\label{fig:process}
	\end{center}
\end{figure}

The relevant signature for this work is represented in Fig.~\ref{fig:process}, which corresponds to first generation squark-pair production followed by the decay into a bino-like $\tilde \chi_3^0$ plus a light jet. Then, one $\tilde \chi_3^0$ decays into the higgsino-like LSP and the SM-like Higgs boson (decaying into $b\bar{b}$). In order to reduce the QCD backgrounds, we consider that the other $\tilde \chi_3^0$ decays into the LSP plus a $Z$ boson decaying into $e^+e^-$ or $\mu^+\mu^-$ pairs.
We study the same supersymmetric spectra as in our previous work~\cite{Arganda:2021ajy}, which is not excluded by the validated analyses of {\tt CheckMATE 2.0.24}~\cite{Dercks:2016npn}.
Within these higgsino-LSP MSSM scenarios, one has BR($\tilde \chi_3^0\to \tilde \chi_{1,2}^0\,h$)$\sim$BR($\tilde \chi_3^0\to \tilde \chi_{2,1}^0\,Z$) but the lower BR($Z\to l^-l^+$)=6.7\% ($l=e,\mu$) with respect to BR($h\to b\bar{b}$)=58\% is compensated by a cleaner final state than in our previous work, yielding to a complementary channel to these spectra.

Concerning the experimental bounds on our MSSM scenarios, we do not have found any LHC search with the same final state and similar spectra. As far as we know, the most related analysis to our study would be Refs.~\cite{CMS:2020bfa,ATLAS:2022zwa}. However, in these works it is assumed that there are only Higgs bosons in the electroweak production and with a massless gravitino. Besides, for colored-particle production no $b$-tagging is done and it is assumed that the decay of the second neutralino is via a $Z$ boson without Higgs bosons. Therefore, our proposed MSSM scenarios evade again such exclusion limits.
In addition, our comparison with {\tt CheckMATE} gives Ref.~\cite{ATLAS:2017vat} as the most sensitive search (but very far from the exclusion) for its signal region $SRI-MLL-60$ when looking for compressed supersymmetric spectra with leptons and missing transverse momentum in the final state (but no $b$-jets).
Finally, supersymmetric signatures with leptons, $b$-jets, and missing energy are produced by top squarks, see for instance~\cite{ATLAS:2016xcm}, thus they are not sensitive to our signature with Higgs and $Z$ bosons.

The relevant backgrounds are $t\bar{t}$+jets and $t\bar{t}$ production in association with a vector boson. The leptons coming from the $Z$-boson decay in the signature reject the presence of the QCD multijet background in this analysis.
We then consider the following SM backgrounds: the fully leptonic decay of $t\bar{t}$ pair up to one additional light jet, $t\bar{t}_{\rm lep}+j$ (inc.); the hadronic decay of $t\bar{t}$ in association to a leptonic $Z$ boson, $t\bar{t}_{\rm had}+Z$; the semileptonic decay of $t\bar{t}$ with a leptonic $Z$ boson, $t\bar{t}_{\rm semilep}+Z$; and the semileptonic decay of $t\bar{t}$ in association with a leptonic $W$ boson, $t\bar{t}_{\rm semilep}+W$.
The jet matching and merging is performed by the MLM algorithm~\cite{Mangano:2002,Mangano:2006rw} using {\tt xqcut}=20 for all generated samples and {\tt qcut} equal to 50 and 250 for backgrounds and signal, respectively.

We demand two $b$-jets and two opposite sign (OS) same flavor leptons (electrons or muons) at reconstructed level:
\begin{equation}
\text{Selection cuts: } N_b = 2 \text{ and } N_\text{lep}^\text{OS} = 2 \,.
\label{eq:selcut}
\end{equation}

\begin{figure}[ht!]
	\begin{center}
		\begin{tabular}{cc}
			\centering
			\hspace*{-3mm}
			\includegraphics[scale=0.44]{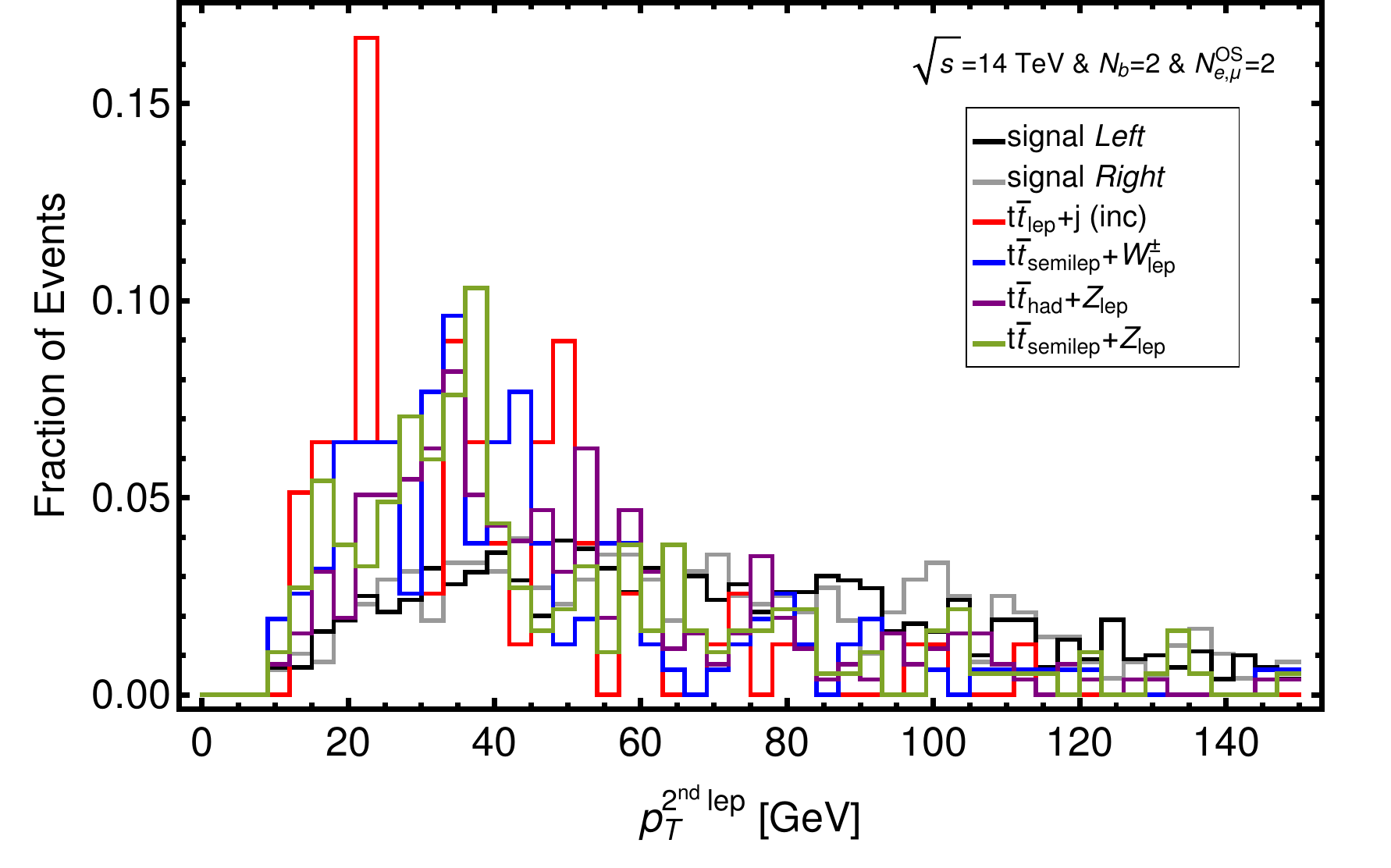} &	\includegraphics[scale=0.44]{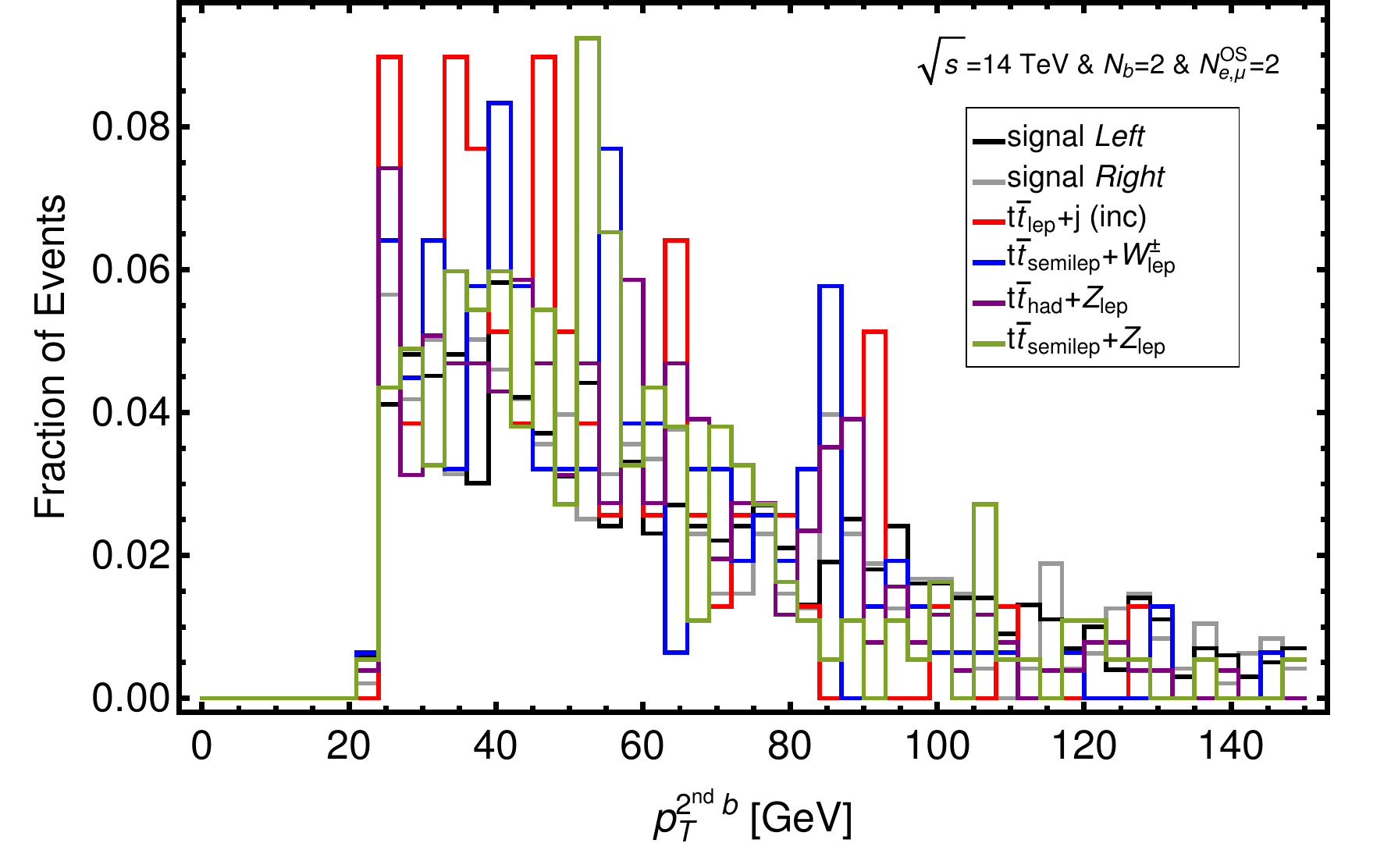}  \\	
		\end{tabular}
		\caption{\it Distributions of transverse momentum of the second leading lepton $p_T^{2^{nd}\,lep}$ (left panel) and second leading $b$-jet $p_T^{2^{nd}\,b}$ (right panel) over samples of signal and backgrounds without parton level cuts.}
		\label{fig:distsincuts1}
	\end{center}
\end{figure}

\begin{figure}[ht!]
	\begin{center}
			\centering
			\hspace*{-3mm}
			\includegraphics[scale=0.7]{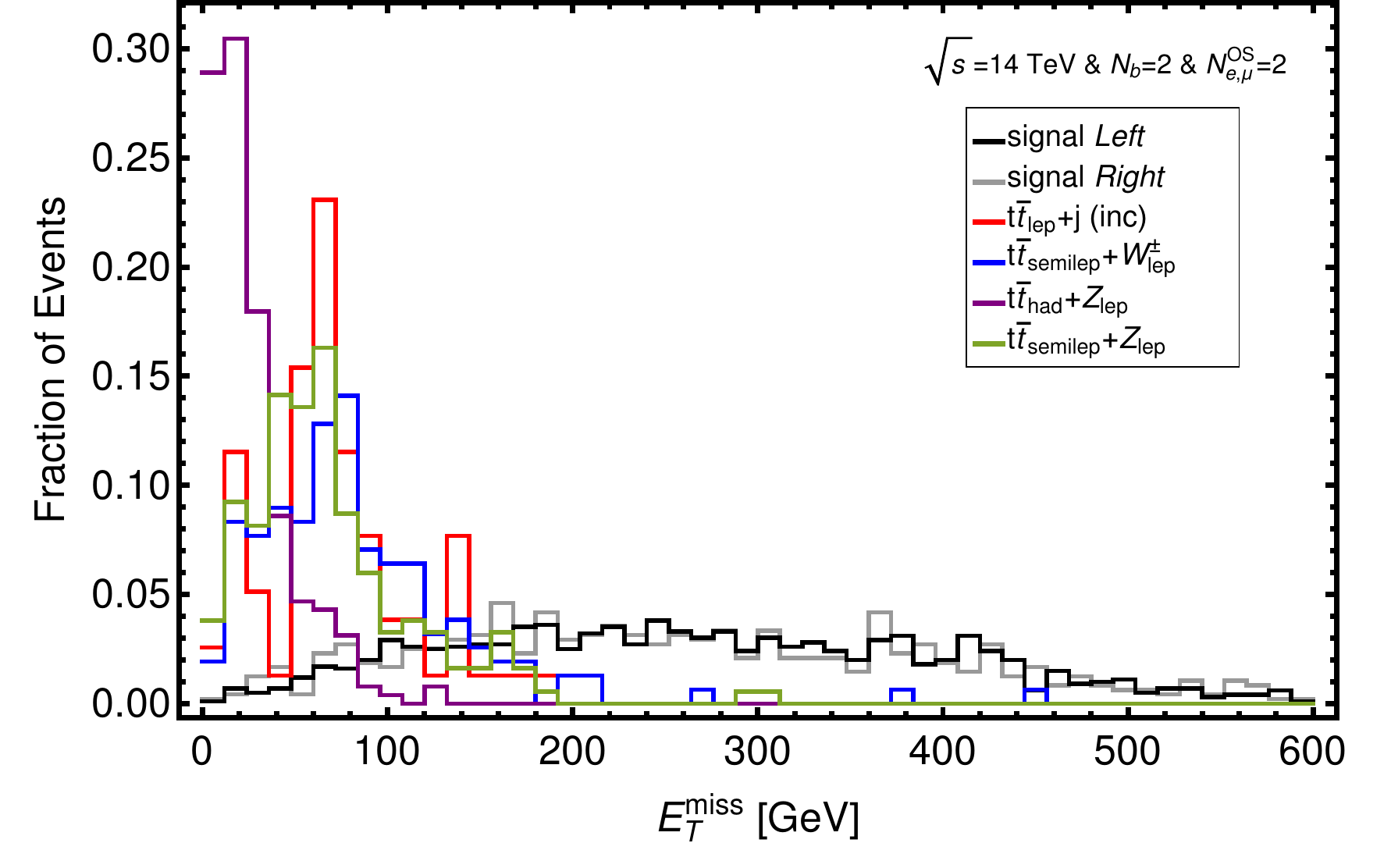} 
		\caption{\it Distribution of missing energy $E_T^\text{miss}$ over samples of signal and backgrounds without parton level cuts.}
		\label{fig:distsincuts2}
	\end{center}
\end{figure}

In order to optimize our background simulation, we display in Fig.~\ref{fig:distsincuts1} the transverse momentum of the second leading lepton $p_T^{2^{nd}\,lep}$ (left panel), the second leading $b$-jet $p_T^{2^{nd}\,b}$ (right panel) and, in Fig.~\ref{fig:distsincuts2}, the missing transverse energy $E_T^\text{miss}$ over samples of signal and backgrounds without parton-level cuts.
We conclude from these three distributions that the parton-level cuts of $p_T^b>25$ GeV, $p_T^{lep}>25$ GeV and $E_T^\text{miss} >$ 200 GeV applied to the background simulation reduce the large cross sections and the event generation becomes more efficient. With this setup, the cross sections of the $t\bar{t}_{\rm lep}+j$ (inc.), $t\bar{t}_{\rm had}+Z$, $t\bar{t}_{\rm semilep}+Z$, and $t\bar{t}_{\rm semilep}+W$ backgrounds are 1.1$\times 10^3$, 19.42, 0.53 and 0.83 fb, respectively. 


\begin{figure}[ht!]
	\begin{center}
		\begin{tabular}{cc}
			\centering
			\hspace*{-3mm}
			\includegraphics[scale=0.44]{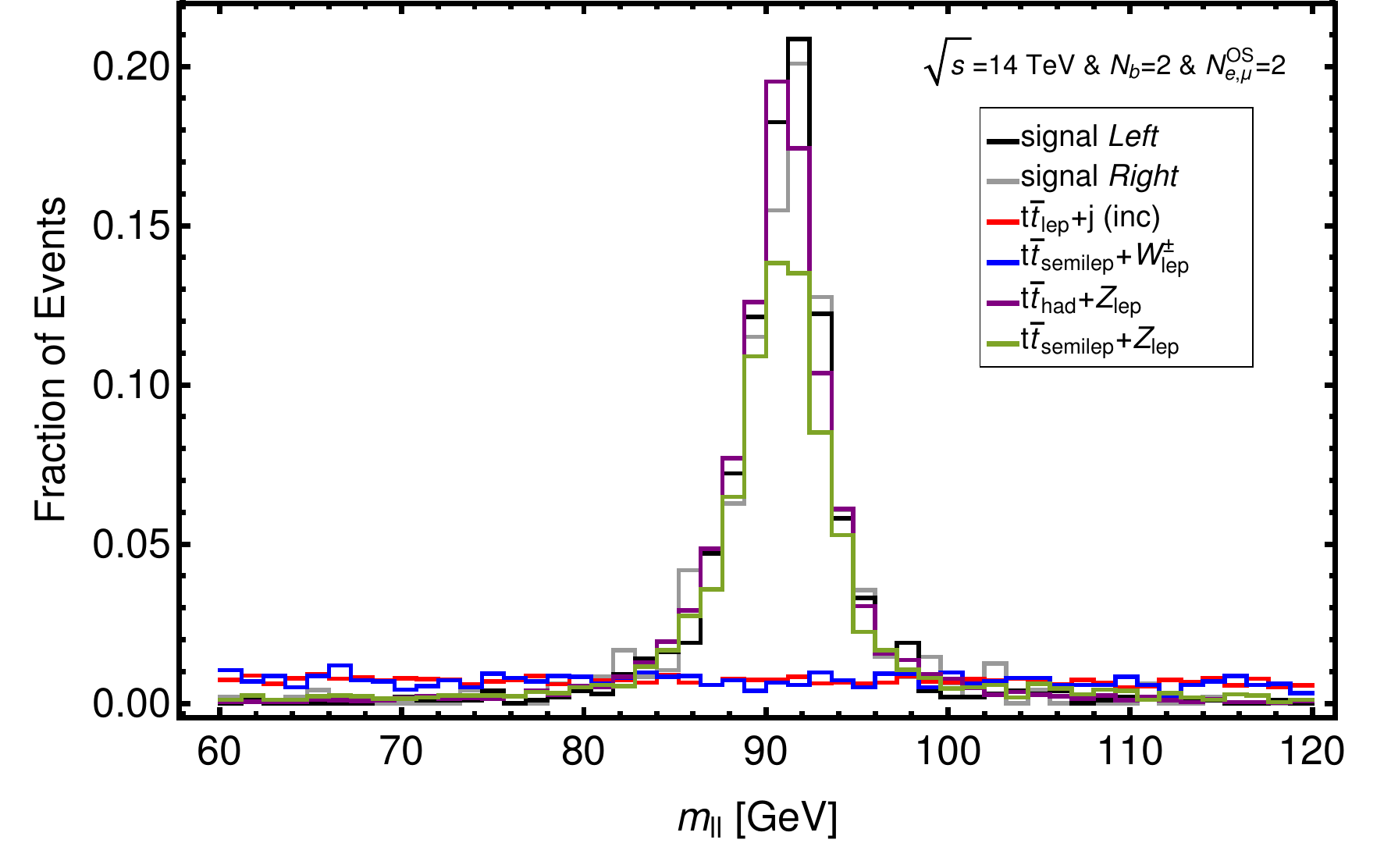} &	\includegraphics[scale=0.44]{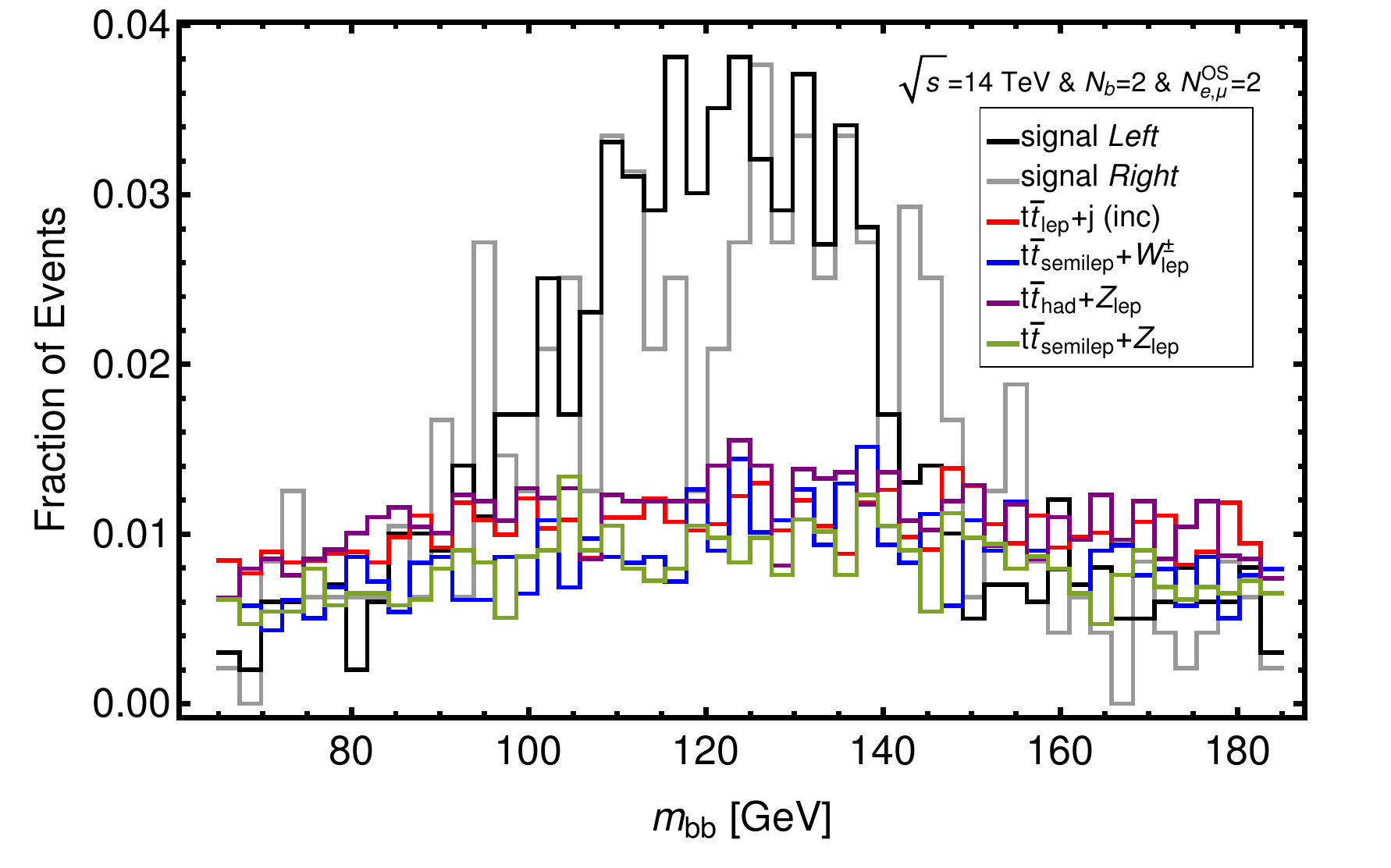}  \\
		\end{tabular}
		\caption{\it Distributions of the invariant mass of a pair of OS leptons $m_{ll}$ (left panel) and a pair of $b$-jets $m_{bb}$ (right panel) over samples of signal and backgrounds without parton level cuts.}
		\label{fig:distsincuts3}
	\end{center}
\end{figure}

On the other hand, the invariant mass of a pair of OS leptons $m_{ll}$ and a pair of $b$-jets $m_{bb}$ are shown in the left and right panels of Fig.~\ref{fig:distsincuts3}. Then it is natural to demand values of these kinematical variables in a 10\% window around the $Z$ and Higgs boson mass, respectively.

Therefore, we applied at detector level the following cuts:
\begin{equation}
p_T^{lep}>25\text{ GeV, }p_T^{b}>25\text{ GeV and }E_T^\text{miss}>200\text{ GeV} \,,
\label{eq:metpt}
\end{equation}
and
\begin{equation}
\left|\frac{m_{ll}-m_Z}{m_Z}\right|<0.1\text{ and }\left|\frac{m_{bb}-m_h}{m_h}\right|<0.1 \,.
\label{eq:minv}
\end{equation}

We develop a search strategy for a luminosity of $\cal{L}$ = 1 ab$^{-1}$, corresponding to the high luminosity LHC phase (HL-LHC), and optimize our analysis for a benchmark with squark masses of 1 TeV for both {\it Left} and {\it Right} productions.
After requiring Eqs.~(\ref{eq:selcut}-\ref{eq:minv}), the $t\bar{t}_{\rm had}+Z$ background disappears. In order to mostly reduce the backgrounds with leptons and missing transverse energy coming from the $W$ boson, we resort to the transverse mass of the second leading lepton $\vec{p}^{\,2^{nd}\,lep}$ and the missing momentum $\vec{p}_T^\text{\,miss}$ given by  
\begin{equation}
m_T^{2^{nd}\,lep}\equiv m_T(\vec{p}^{\,2^{nd}\,lep},\vec{p}_T^\text{\,miss})=\sqrt{2p_T^{\,2^{nd}\,lep}E_T^\text{miss}(1-\cos{\Delta\phi(\vec{p}^{\,2^{nd}\,lep},\vec{p}_T^\text{\,miss})})} \,.
\end{equation}

\begin{figure}[ht!]
	\begin{center}
		\begin{tabular}{cc}
			\centering
			\hspace*{-3mm}
			\includegraphics[scale=0.44]{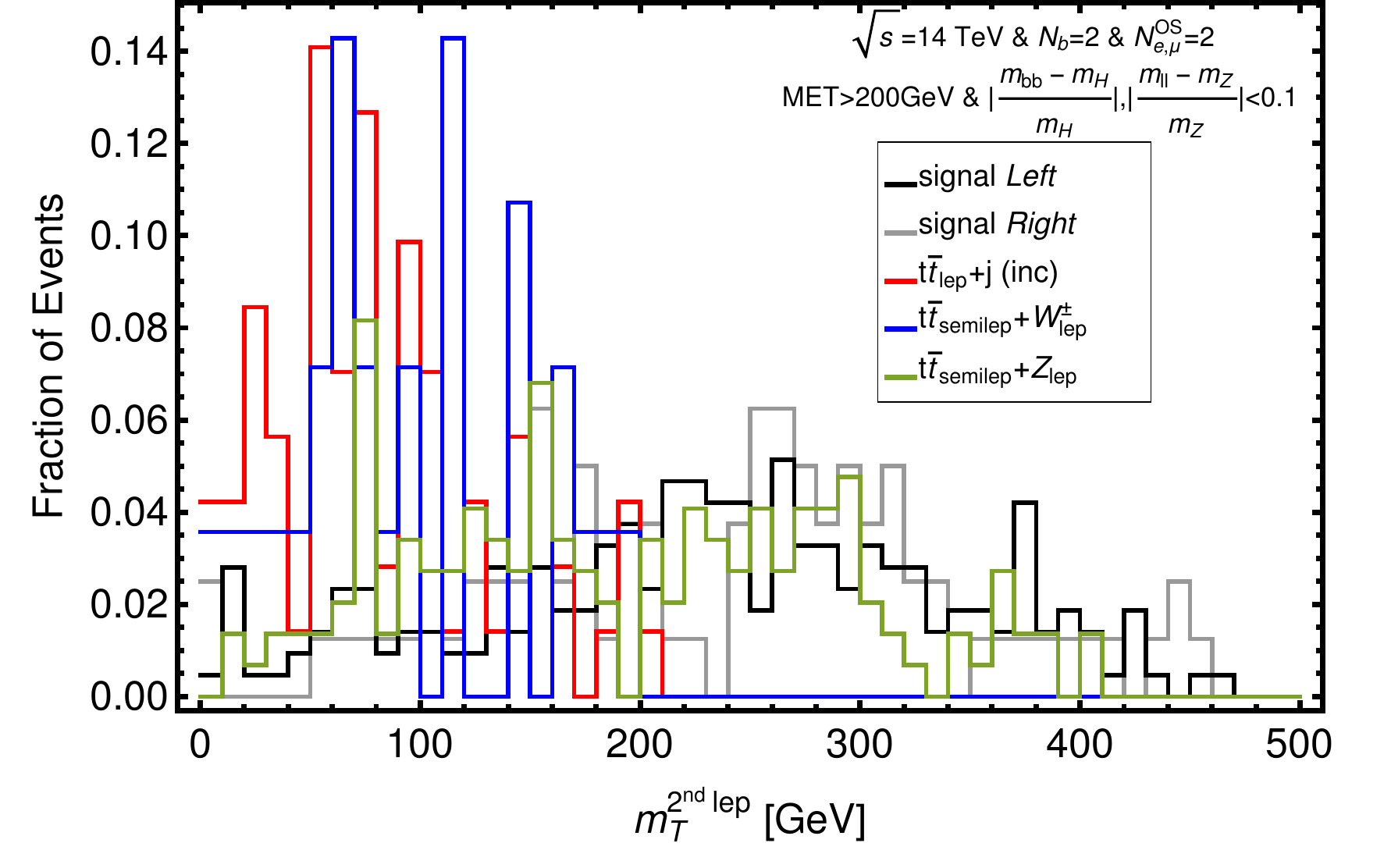} &			\includegraphics[scale=0.44]{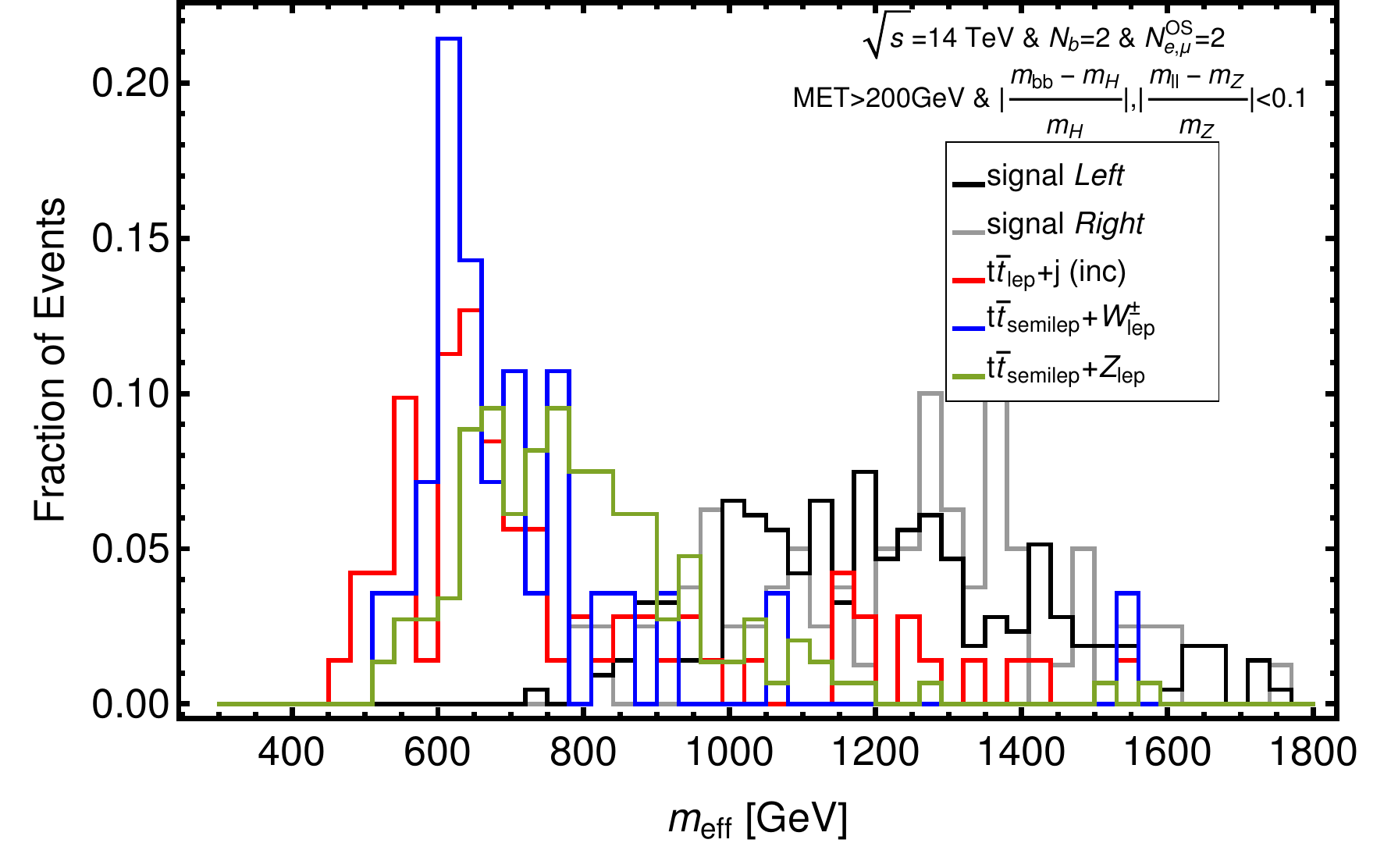} \\
		\end{tabular}
		\caption{\it Distributions of the transverse mass of the second lepton $m_T^{2^{nd}\,lep}$ (left) and the effective mass $m_\text{eff}$ (right), for signal and background after requiring two $b$-jets, two OS leptons and satisfying Eqs.~(\ref{eq:selcut}-\ref{eq:minv}).}
		\label{fig:distfinales}
	\end{center}
\end{figure}

We also consider the effective mass variable $m_\text{eff}$ corresponding to the scalar sum of the missing energy and the transverse momentum of all reconstructed objects.
Figure~\ref{fig:distfinales} shows the distributions of these two variables after demand the cuts of Eqs.~(\ref{eq:selcut}-\ref{eq:minv}).
We can see from the left panel of Fig.~\ref{fig:distfinales} that the $t\bar{t}_{\rm lep}+j$ (inc.) and $t\bar{t}_{\rm semilep}+W^\pm$ backgrounds have values below 220 GeV for this variable. From the effective mass distributions, we found that background peaks are below 1000 GeV while the signal have most of its events above this value.

In order to obtain an estimate of the LHC sensitivity to our SUSY signal, we use the following expression for the signal significance, including background systematic uncertainties~\cite{Cowan:2010js,Cowan:2012}:
\begin{equation}
{\cal S} = \sqrt{2 \left((B+S) \log \left(\frac{(S+B)(B+\sigma_{B}^{2})}{B^{2}+(S+B)\sigma_{B}^{2}}\right)-\frac{B^{2}}{\sigma_{B}^{2}}\log \left(1+\frac{\sigma_{B}^{2}S}{B(B+\sigma_{B}^{2})} \right) \right)} \,,
\label{systS}
\end{equation}
where $S$ ($B$) is the number of signal (background) events and $\sigma_{B}=(\Delta B) B$, with $\Delta B$ being the relative systematic uncertainty chosen to be a conservative value of 30\%. 


\section{Discussion and Summary}
\label{results}

\begin{table}
\hspace*{-12.5mm}
    \centering
\begin{tabular}{r|rrrrr|cc}
\hline\hline
   Process  & signal & $t\bar{t}_{\rm lep}+j$ (inc.) & $t\bar{t}_{\rm had}+Z$ & $t\bar{t}_{\rm semilep}+Z$ & $t\bar{t}_{\rm semilep}+W^\pm$ & $\cal{S}$ \\
   \hline
    Expected  & 239 & $1.11 \times 10^6$ & 19420 & 530 & 830 & $7\times 10^{-4}$ \\
    \hline
    selection cut  & 29.8 & 40888 & 2140 & 38 & 39 & $2.3\times 10^{-3}$ \\
    cuts of Eqs.~(\ref{eq:metpt}-\ref{eq:minv})  & 5.9 & 200 & 0 & 1.6 & 0.27 & 0.09 \\
    $m_T^{2^{nd}lep}>220$ GeV & 3.7 & 0 & 0 & 0.86 & 0 & 2.61 \\
    $m_{eff}>1000$ GeV & 3.4 & 0 & 0 & 0.11 & 0 & 4.02 \\
    \hline\hline
    $\cal{L}$ = 3 ab$^{-1}$& 10.2 & 0 & 0 & 0.33 & 0 & 6.65 \\
    \hline\hline
\end{tabular}
    \caption{\it Cut flow with $\cal{L}$ = 1 ab$^{-1}$ corresponding to the \textit{Left} case signal production. The last column is the significance with a systematic uncertainty in the background of 30\%. The last row is the projection to $\cal{L}$ = 3 ab$^{-1}$.}
    \label{cutflowLEFT}
\end{table}

\begin{table}
\hspace*{-12.5mm}
    \centering
\begin{tabular}{r|rrrrr|cc}
\hline\hline
   Process  & signal & $t\bar{t}_{\rm lep}+j$ (inc.) & $t\bar{t}_{\rm had}+Z$ & $t\bar{t}_{\rm semilep}+Z$ & $t\bar{t}_{\rm semilep}+W^\pm$ & $\cal{S}$ \\
   \hline
    Expected  & 195 & $1.11 \times 10^6$ & 19420 & 530 & 830 & $5.7\times 10^{-4}$ \\
    \hline
    selection cut  & 23.4 & 40888 & 2140 & 38 & 39 & $1.8\times 10^{-3}$ \\
   cuts of Eqs.~(\ref{eq:metpt}-\ref{eq:minv})  & 3.5 & 200 & 0 & 1.6 & 0.27 & 0.06 \\
    $m_T^{2^{nd}lep}>220$ GeV & 2.3 & 0 & 0 & 0.86 & 0 & 1.76 \\
    $m_{eff}>1000$ GeV & 1.76 & 0 & 0 & 0.11 & 0 & 2.61 \\
    \hline\hline
    $\cal{L}$ = 3 ab$^{-1}$& 5.3 & 0 & 0 & 0.33 & 0 & 4.37 \\
    \hline\hline
\end{tabular}
    \caption{\it Cut flow with $\cal{L}$ = 1 ab$^{-1}$ corresponding to the \textit{Right} case signal production. The last column is the significance with a systematic uncertainty in the background of 30\%. The last row is the projection to $\cal{L}$ = 3 ab$^{-1}$.}
    \label{cutflowRIGHT}
\end{table}

The resulting cut flows for both {\it Left} and {\it Right} signal cases are presented in Table~\ref{cutflowLEFT} and \ref{cutflowRIGHT}, respectively, in which the different cuts that define our search strategy are listed. First we can observe that the selection cuts are very useful to drastically reduce the $t\bar{t}_{\rm lep}+j$ background, which is potentially the most problematic, and whose number of events drops by two orders of magnitude after these cuts, while the number of events of the rest of the backgrounds and signal are reduced by about one order of magnitude. The cuts of Eqs.~(\ref{eq:metpt}-\ref{eq:minv}) remove all $t\bar{t}_{\rm had}+Z$ background events and reduce two orders of magnitude of the rest, whilst the signal events decrease less than one order of magnitude. After the $m_T^{2^{nd}lep}$ cut, only the $t\bar{t}_{\rm semilep}+Z$ background survives and the signal is hardly affected. Finally, the $m_{eff}$ cut helps to further reduce the $t\bar{t}_{\rm semilep}+Z$ background with little change in the number of final signal events. At the end, for a total integrated luminosity of $\cal{L}$ = 1 ab$^{-1}$, we expect signal significances of 4.02$\sigma$ and 2.61$\sigma$ for the {\it Left} and {\it Right} cases, respectively. If we project these results for a luminosity of 3 ab$^{-1}$, the significances would reach values of 6.65$\sigma$ and 4.37$\sigma$, respectively, that one can consider at the discovery level of sensitivity.

\begin{figure}
	\begin{center}
		\begin{tabular}{cc}
			\centering
			\hspace*{-3mm}
			\includegraphics[scale=0.9]{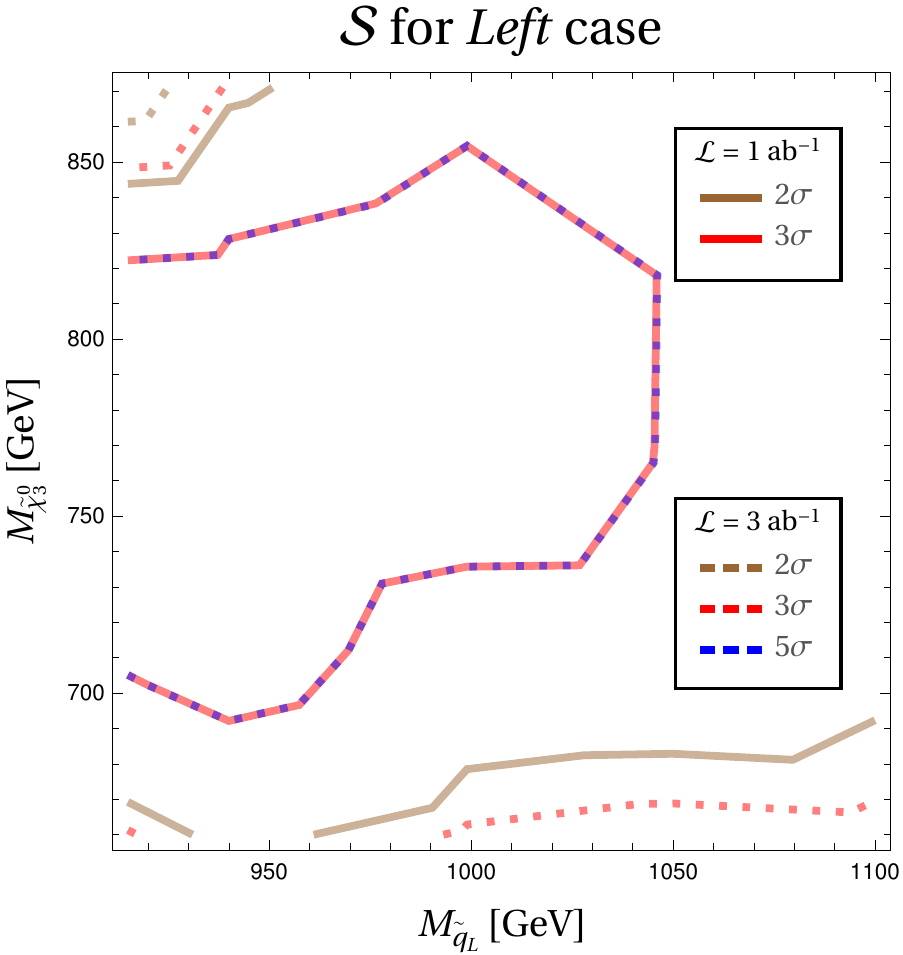} &
			\includegraphics[scale=0.9]{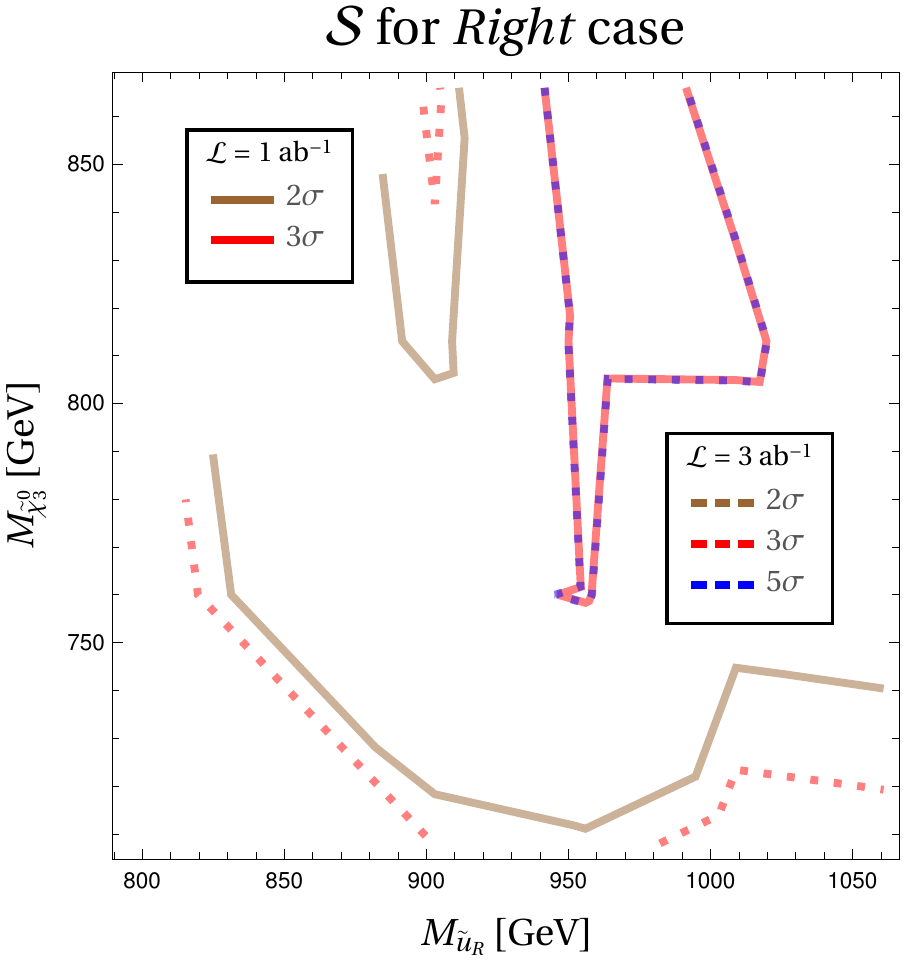}	
		\end{tabular}
		\caption{\it Contour lines in the plane $[M_{\tilde q},M_{\tilde\chi_3^0}]$ for $\tilde{q}_L$ (left) and $\tilde{u}_R$ (right) productions. Solid (dotted) lines correspond to ${\cal L}$ = 1 (3) ab$^{-1}$. The brown, red and blue colors are the $\cal{S}$ (background systematic uncertainty of 30\%) with values of 2$\sigma$, 3$\sigma$ and 5$\sigma$, respectively.}
		\label{contourplots}
	\end{center}
\end{figure}

The promising results in Tables~\ref{cutflowLEFT} and~\ref{cutflowRIGHT}, for the higgsino-LSP MSSM benchmarks with squark masses of 1 TeV, encourage the extension of our analysis to other values of the parameter space of interest, defined in the plane [$M_{\tilde q}$, $M_{\tilde\chi_3^0}$].

We show in Fig.~\ref{contourplots} the contour lines in the plane $[M_{\tilde q},M_{\tilde\chi_3^0}]$ for $\tilde{q}_L$ (left panel) and $\tilde{u}_R$ (right panel) pair productions. The solid (dotted) lines correspond to a luminosity of ${\cal L}$ = 1 (3) ab$^{-1}$. The brown, red, and blue colors represent the values of 2$\sigma$, 3$\sigma$ and 5$\sigma$, respectively, for the signal significance, $\cal{S}$. For the lowest luminosity of 1 ab$^{-1}$, we are able to obtain 2$\sigma$ significances, in the {\it Left} case, for virtually any $M_{\tilde q_L}$ value within the range considered [850 GeV - 1100 GeV] and bino mass values above $\sim$ 650 GeV and below $\sim$ 850 GeV. One would reach significances at the evidence level for values of $M_{\tilde q_L} <$ 1050 GeV and $M_{\tilde\chi_3^0}$ between 700 GeV and 850 GeV. This same area defines the discovery-level sensitivity for ${\cal L}$ = 3 ab$^{-1}$ in the {\it Left} case. Our search strategy, applied to the {\it Right} case, allows to get 2$\sigma$ significances for $M_{\tilde u_R} \gtrsim$ 825 GeV and practically any value of $M_{\tilde\chi_3^0}$ for ${\cal L}$ = 1 ab$^{-1}$. Evidence-level significances are obtained in this case for $M_{\tilde\chi_3^0} \gtrsim$ 800 GeV and $M_{\tilde u_R}$ values between 950 GeV and 1000 GeV. These squark and bino mass values also delimit the discovery-level area in the {\it Right} case for ${\cal L}$ = 3 ab$^{-1}$.

To summarize, we have developed a search strategy for pairs of squarks at the HL-LHC, which both decay into bino neutralinos plus a jet. In turn, one of the binos decays into a Higgs boson plus the LSP, while the other decays into the LSP and a leptonic $Z$ boson, which allows us to keep all backgrounds under control and to smoothly discard the QCD multijet background. Our collider analysis provides signal significances at the evidence level for luminosities of 1 ab$^{-1}$ and at the discovery level if we project these results for 3 ab$^{-1}$.

\section*{Acknowledgments}
The work of EA is partially supported by the ``Atracci\'on de Talento'' program (Modalidad 1) of the Comunidad de Madrid (Spain) under the grant number 2019-T1/TIC-14019, by the Spanish Research Agency (Agencia Estatal de Investigaci\'on) through the Grant IFT Centro de Excelencia Severo Ochoa No CEX2020-001007-S, funded by MCIN/AEI/10.13039/501100011033, and by CONICET and ANPCyT (Argentina) under projects PICT 2017-2751 and PICT 2018-03682. The work of AD was partially supported by the National Science Foundation under grant PHY-2112540. The work of RM is supported by CONICET and ANPCyT under projects PICT 2016-0164, PICT 2017-2751 and PICT 2018-03682. The work of MQ is partly supported by Spanish MINEICO under Grant FPA2017-88915-P, and by the Catalan Government under Grant 2017SGR1069. 
IFAE is partially funded by the CERCA program of the Generalitat de Catalunya.

\bibliographystyle{JHEP}
\bibliography{lit.bib}

\end{document}